\documentclass[twocolumn,prb,showpacs,preprintnumbers,amsmath,amssymb]{revtex4-1}
\usepackage{graphicx}
\usepackage{dcolumn}
\usepackage{bm}
\usepackage{tabularx}
\usepackage{multirow}
\usepackage{color}

\begin{document}

\title {Two different topological insulator phases in NaBaBi}

\author{Yan Sun$^1$}
\author{Qing-Ze Wang$^3$}
\author{Shu-Chun Wu$^1$}
\author{Claudia Felser$^1$}
\author{Chao-Xing Liu$^3$}
\author{Binghai Yan$^{1,2}$}
\email[Corresponding author:]{yan@cpfs.mpg.de}

\affiliation{$^1$ Max Planck Institute for Chemical Physics of Solids, 01187 Dresden, Germany}
\affiliation{$^2$ Max Planck Institute for the Physics of Complex Systems, 01187 Dresden,Germany}
\affiliation{$^3$ Department of Physics, The Pennsylvania State University, University Park, Pennsylvania 16802-6300}

\date{\today}

\begin{abstract}
By breaking the inversion symmetry of the three-dimensional Dirac metal Na$_3$Bi,
we realize topological insulator (TI) phases in the known compound
NaBaBi using $ab~initio$ calculations. Two distinct  TI phases
emerge: one phase is due to band inversion between the Bi $p$ and
Na $s$ bands, and the other phase (under pressure) is induced by
inversion of the Bi $p$ and Ba $d$ bands. Both phases exhibit
Dirac-cone-type surface states, but they have opposite spin textures. In
the upper cone, a left-handed spin texture exists for the $s$-$p$
inverted phase (similar to a common TI, e.g., Bi$_2$Se$_3$),
whereas a right-handed spin texture appears for the $p$-$d$ inverted
phase. NaBaBi presents a prototype model of a TI that exhibits
different spin textures in the same material. In addition, NaBaBi
may exhibit the introduction of correlation effects to the topological state
owing to the existence of $d$ states in the band inversion.
\end{abstract}


\maketitle

\section{introduction}
In recent years, topological insulators (TIs) have attracted
considerable attention owing to their novel fundamental physics
and potential applications.~\cite{hasan,xiao}
Because of the nontrivial topology of the bulk band structure,
TIs exhibit gapless surface states inside the bulk energy gap.
The spin direction is locked to the momentum by spin--orbit coupling (SOC). Thus, the topological surface state (TSS)
exhibits a helical spin texture in momentum space.
To date, left-handed spin texture is commonly found in known TI materials
such as the Bi$_2$Se$_3$ family~\cite{BiSe4,BiSe3,BiSe1,BiSe2,Liu2010} and other materials,\cite{BBO3,Na3Bi,BiTlSe2}
whereas right-handed spin texture has been reported only in HgS.~\cite{HgS}

Topological Dirac semimetals were discovered recently in Na$_3$Bi ~\cite{Na3Bi,Na3Bi-exp} and Cd$_3$As$_2$.~\cite{Wang2013,Liu2014,Neupane2014,Jeon2014,Yi2014} They also exhibit nontrivial topology in the band structure and can be driven into
other exotic topological phases, such as TIs and Weyl metals.~\cite{Wan2011}
For example, the three-dimensional (3D) Dirac point can become massive by breaking a certain crystal symmetry, inducing a TI phase.
Following this paradigm, we report a TI material, NaBaBi, synthesized in a recent experiment,~\cite{NaBaBi}  which is 
equivalent to breaking the inversion symmetry of Na$_3$Bi by replacing two Na sites with one Ba.
Furthermore,  NaBaBi can be transformed between two different TI phases via external pressure,
one with $s$-$p$ band inversion and left-handed spin texture and the other with $p$-$d$ band inversion and right-handed spin texture.

\begin{figure}[htbp]
\begin{center}
\includegraphics[width=0.45\textwidth]{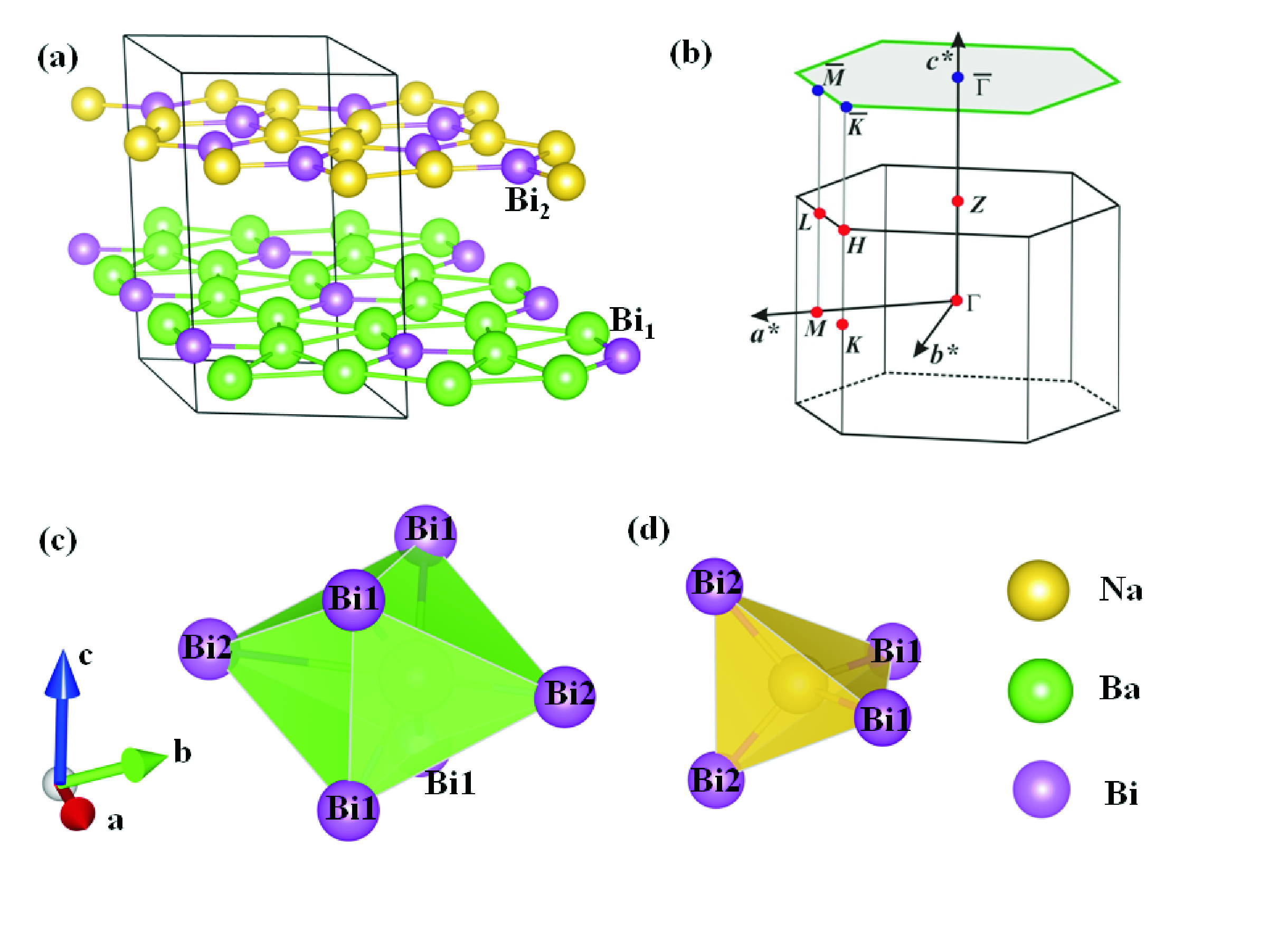}
\end{center}
\caption{
(color online) (a) Hexagonal crystal lattice for NaBaBi. 
Two atomic layers, [(Bi$^{3-})_{2}$(Na$^{+})_{3}]^{3-}$ and [(Bi$^{3-}$)(Ba$^{2+})_{3}]^{3+}$,
are stacked alternately along the $c$ direction. 
(b) Corresponding hexagonal Brillouin zone (BZ) and two-dimensional (2D) BZ projected onto the (001)
surface. (c), (d) Local environments around Ba and Na atoms are 
octahedra and tetrahedra, respectively, distorted by Bi atoms.
}
\label{crystal}
\end{figure}

\begin{figure}[htbp]
\begin{center}
\includegraphics[width=0.45\textwidth]{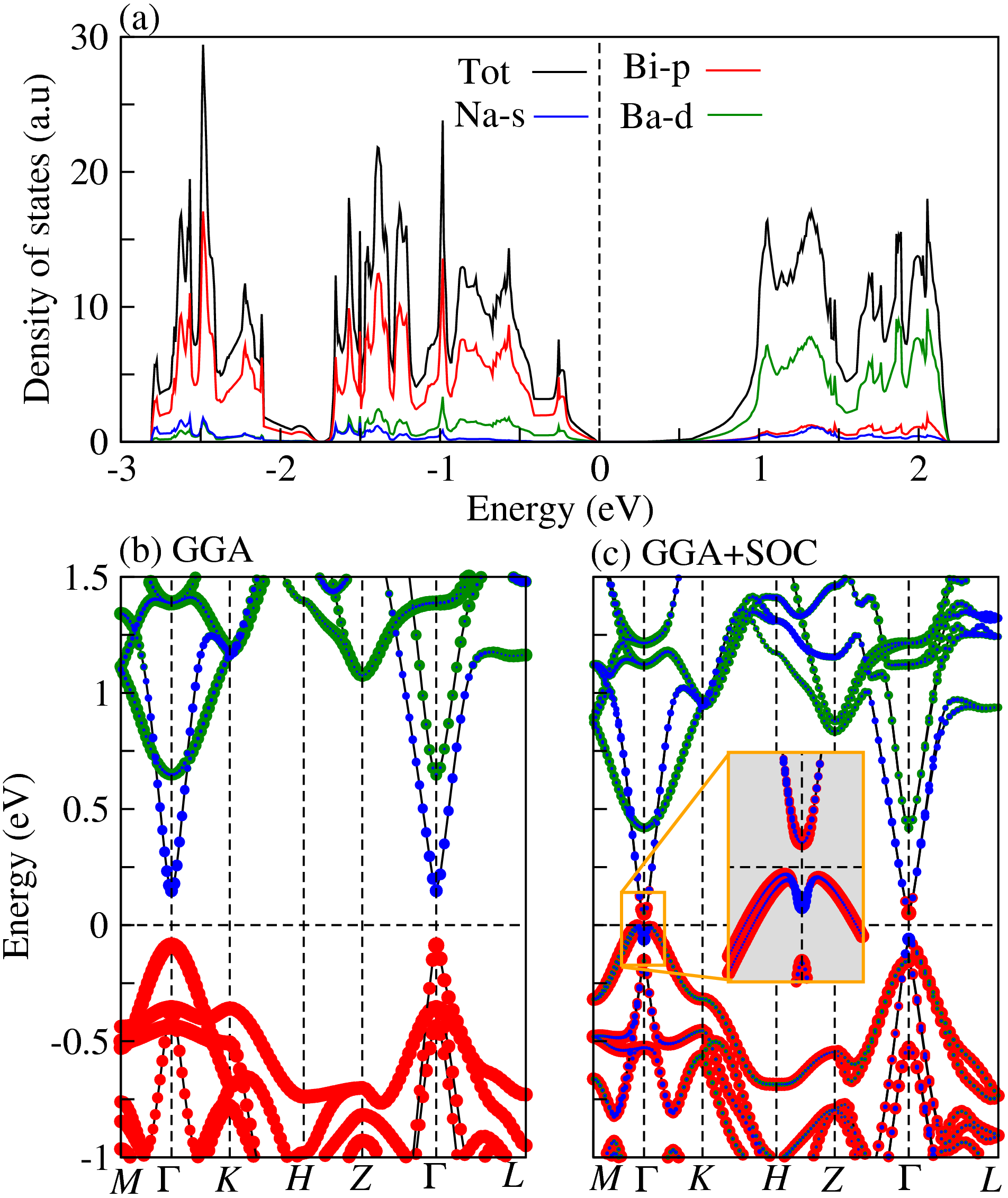}
\end{center}
\caption{(color online)
Electronic properties of NaBaBi.
(a) Total and partial density of states.
(b,c) Energy dispersions along high-symmetry lines in lattice
momentum space for NaBaBi without and with SOC. 
Sizes of blue, red, and green dots in (b) and (c)
are proportional to the weight contributions of Na $s$, Bi $p$, and Ba $d$ orbitals, respectively. 
Fermi energy has been shifted to zero.
}\label{band_structure}
\end{figure}

\section{Crystallographic structure and calculation methods}

NaBaBi was first synthesized via direct reaction of the elements Na,
Ba, and Bi at high temperature in 2004.~\cite{NaBaBi} As shown in Fig. 1(a), NaBaBi exhibits
a hexagonal crystal lattice in the $P\overline{6}2m$ space group (No. 189).
 It is a polar intermetallic compound:
two atomic layers are stacked along the $c$ direction in each unit cell, one layer with [(Bi$^{3-})_{2}$(Na$^{+})_{3}]^{3-}$ and the other with [(Bi$^{3-}$)(Ba$^{2+})_{3}]^{3+}$. 
We can understand the lattice of NaBaBi as a
distorted version of that of Na$_3$Bi. Two equivalent Bi sites that are connected by the inversion symmetry in Na$_3$Bi
become non-equivalent in NaBaBi. The surrounding Ba and Na atoms are in
octahedral and tetrahedral environments that are distorted by Bi atoms, as shown in Figs. 1(c) and (d), respectively.
In principle, NaBaBi can also be regarded as a low-symmetry version of layered honeycomb
ternary compounds, which were recently found to be strong~\cite{Zhang2011} and weak TIs.~\cite{Yan2012}
Because the number of total valence electrons is eight in a formula unit, the closed-shell configuration
indicates that NaBaBi is possibly an insulator.

\begin{figure} [htpb]
\begin{center}
\includegraphics[width=0.45\textwidth]{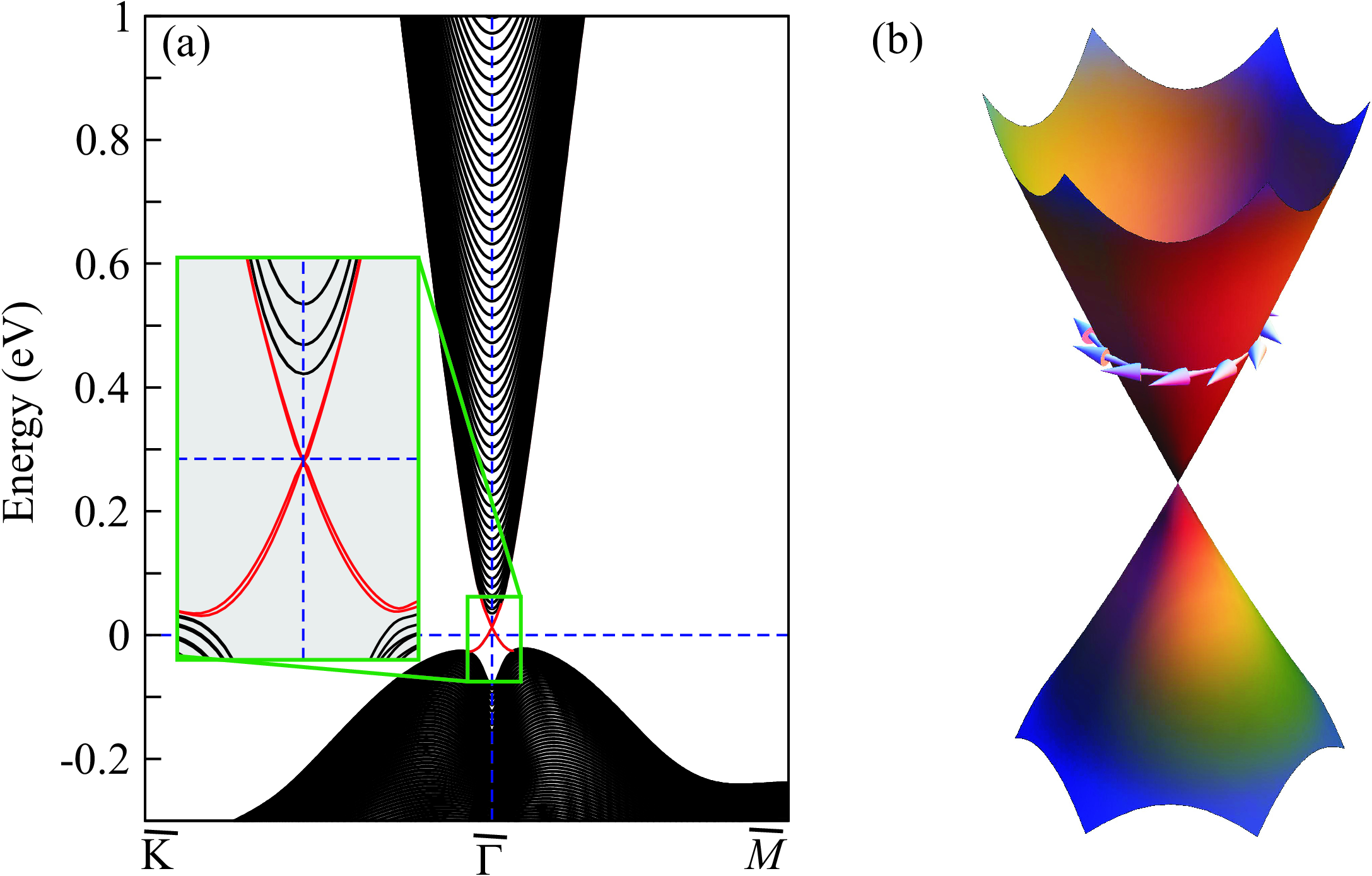}
\end{center}
\caption{(color online)
TB energy band structures of the (a) (001) surface and (b) corresponding left-handed spin
texture around the Fermi energy for NaBaBi. Inset shows Dirac points resembling local energy 
dispersion around the Fermi energy at the $\Gamma$ points. The surface
projected momenta are indicated in Fig. 1(b). Fermi energy is set to zero.
} \label{surface}
\end{figure}

To investigate the band structure of NaBaBi, we performed
electronic structure calculations using density functional theory (DFT)
with the projected augmented wave method,~\cite{PAW} as implemented in the 
Vienna $ab~initio$ Simulation Package.~\cite{Kresse, Kresse2} The
exchange-correlation energy was considered in the generalized gradient
approximation (GGA) level with the Perdew--Burke--Ernzerhof-based density
functional.~\cite{pbe} The energy cutoff is set to 400 eV for the plane wave basis. 
We adopted k-point grids of $5\times5\times7$ and $7\times7\times9$ for the lattice optimization and
electronic structure calculations, respectively.
The surface band structures are calculated on a slab model in a tight-binding (TB) scheme based on the maximally localized Wannier functions (MLWFs),~\cite{mostofi} 
 which are projected from the bulk Bloch wave functions.

\section{Results and discussion}

The band structures in Fig. 2 show that bands
near the Fermi energy are contributed by the Bi $p$, Na $s$, and Ba $d$
orbitals. Without SOC, NaBaBi is a direct-gap semiconductor with a
band gap of 0.23 eV at the $\Gamma$ point. The highest occupied band is a 
Bi $p_z$ state, and the lowest unoccupied band is a Na $s$ state, as shown in Fig. 2(b).
When the SOC effect is included, the band structure around the $\Gamma$ point changes
dramatically. 
SOC pushes the Bi $p_z$ band above the Na $s$ band,
producing an inverted band structure with an M-shaped valence band [Fig. 2(c)].
Although the $Z_2$ index cannot be directly calculated from the parity eigenvalues of the Bloch wave functions~\cite{Fu2007}
 owing to the inversion asymmetry, the SOC-induced band inversion clearly exhibits
a topological phase transition from a trivial insulator to a TI.

The TI phase is further confirmed by the direct surface
state calculations. 
Considering that band inversion occurs between the Na $s$ and Bi $p$ states 
near the Fermi energy,
the MLWFs are derived from the atomic-like Na $s$ and Bi $p$ orbitals to obtain the TB parameters.
Then we construct the Hamiltonians of a slab for the (001) surface with a thickness of
200 unit cells and obtain the surface band structure from direct diagonalization.
As shown in Fig. 3,  the surface exhibits nontrivial
Dirac-cone-type TSSs inside the bulk band gap. 
Because of the absence of inversion symmetry,  
the two Dirac cones on the top and bottom surfaces of the slab are not degenerate in energy,
 especially when they extend far from the Brillouin zone (BZ) center.
As in most TIs, the surface Dirac cone of NaBaBi exhibits left-handed spin texture in the upper Dirac cone [Fig. 3(b)].
These results unambiguously demonstrate the topologically nontrivial features of NaBaBi.

As observed in the bulk band structure, 
the Ba $5d$ state appears as the second-lowest conduction band above the Na $s$ state. 
Next, we show that the compressive pressure can shift the 
Ba $d$ band down to realize an inversion of the Ba $d$ and Bi $p$ bands, i.e., a new TI phase.
At the same time, the Na $s$ state is pushed up into the conduction bands, and the original 
Na $s$ and Bi $p$ inversion is removed.
In the crystal field of the BaBi$_6$ octahedron [Fig. 1(c)],
the low-energy $t_{2g}$ bands are further split by strong distortion of the octahedron,
where Ba $d_{z^2}$ becomes the lowest band of the $t_{2g}$ states 
(the $z$ direction corresponds to the crystal $c$ axis).
Under compressive pressure that shrinks the BaBi$_6$ octahedron, 
the crystal field splitting of the Ba $d$ states increases, so 
the lowest band, Ba $d_{z^2}$, is pushed down in energy. 
In the crystal field of the NaBi$_4$ tetrahedron [Fig. 1(d)], 
compressive pressure will simply increase the energy of the Na $s$ state 
by enhancing the Coulomb repulsion of Bi$^{3-}$ anions.

\begin{figure}[htbp!]
\begin{center}
\includegraphics[width=0.45\textwidth]{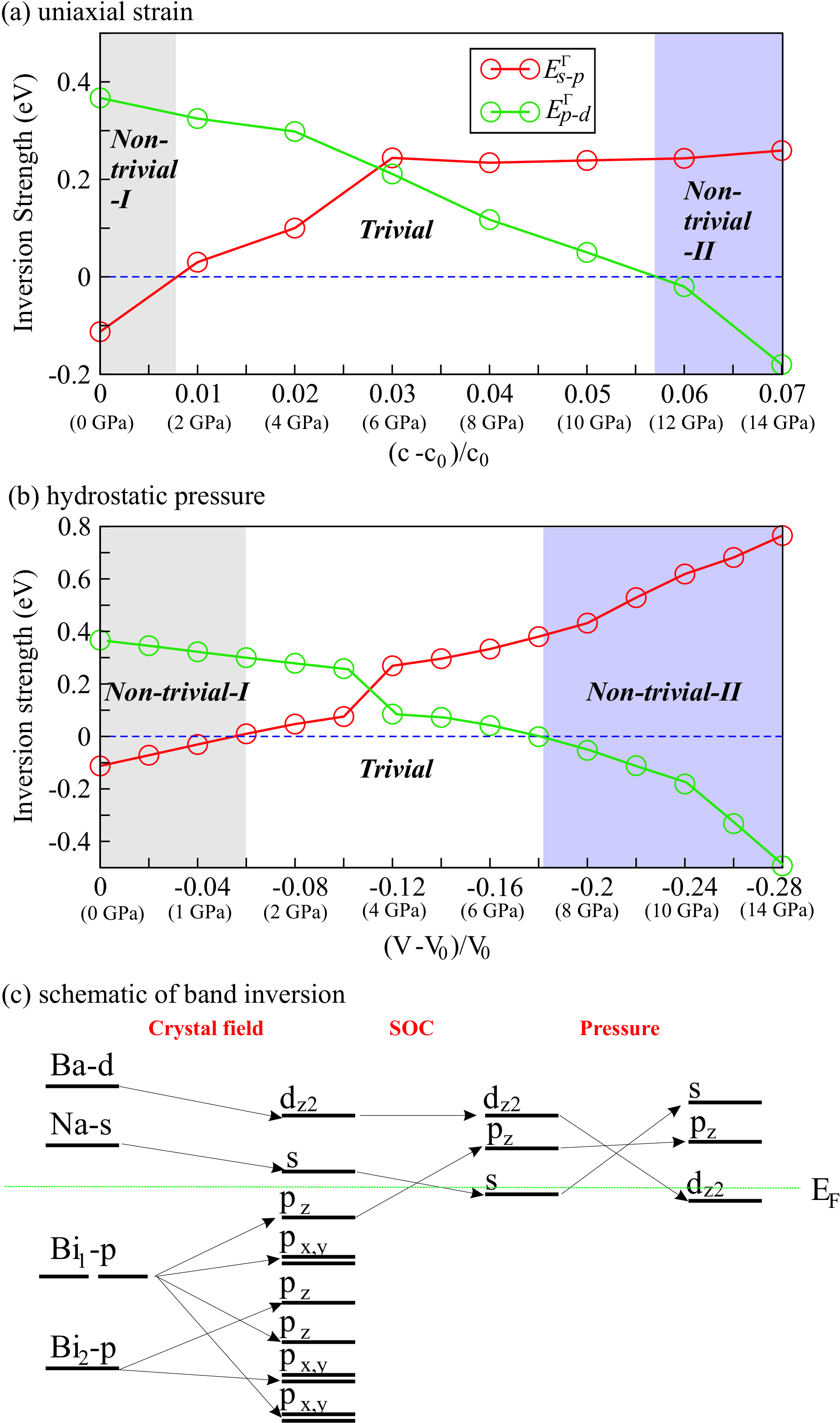}
\end{center}
\caption{(color online)
Evolution of band inversion strengths of
$E_{s-p}^{\Gamma}=E_{Na-s}^{\Gamma}-E_{Bi-p_{z}}^{\Gamma}$ and
$E_{d-p}^{\Gamma}=E_{Ba-d_{z2}}^{\Gamma}-E_{Bi-p_{z}}^{\Gamma}$
under (a) uniaxial strain in $c$ direction and (b) hydrostatic
pressure.
Negative values of the inversion strength represent inverted band gaps, i.e., the TI phase.
$V_0$ and $c_0$ are the volume
and lattice constant in the equilibrium state, respectively, whereas $V$ and $c$ are those under
the pressure or strain state.  The two shaded areas
in both top and bottom panels are topologically nontrivial states
with $p$-$d$ (nontrivial-II) and $s$-$p$ band inversion (nontrivial-I),
and the bright blocks are topologically trivial states.
} \label{energy-differ}
\end{figure}

In the calculations, we apply the external pressure in two different ways:
as hydrostatic pressure and as uniaxial pressure.
Under hydrostatic pressure, we first estimate the bulk modulus by fitting
 the total energy dependence on the volume with
 Murnaghan's equation ~\cite{Murnaghan} and obtained
 $B_0=25$ GPa, which awaits validation by future experiments.
 Under uniaxial pressure, a tensile strain is applied
along the $c$ axis while the $a$ and $b$ axes are optimized, 
which is equivalent to applying in-plane compressive strain in the $ab$ plane, 
such as that induced by a substrate.
Under both types of pressure, the Ba $d_{z^2}$ band shifts down and
 the Na $s$ band shifts up, as expected.
As a consequence, two topological phase transitions occur with increasing pressure.
During the first transition, at a hydrostatic pressure of 1.5 GPa (uniaxial pressure of 2 GPa),
because of the upshift of the Na $s$ state, the band inversion
between Na $s$ and Bi $p$ disappears. Thus, NaBaBi changes from a TI into a normal insulator.
During the second transition, at a hydrostatic pressure of 7 GPa (uniaxial pressure of 12 GPa), because of the downshift of the Ba $d_{xy}$ state,
a new inversion occurs between Ba $d$ and Bi $p$ at the $\Gamma$ point.
Therefore, NaBaBi changes from a normal insulator to a TI phase.
We show the phase diagram with respect to the pressure in Fig. 4, where two TI phases, nontrivial-I
and nontrivial-II, are separated by a trivial insulator phase.

\begin{figure} [htpb!]
\begin{center}
\includegraphics[width=0.45\textwidth]{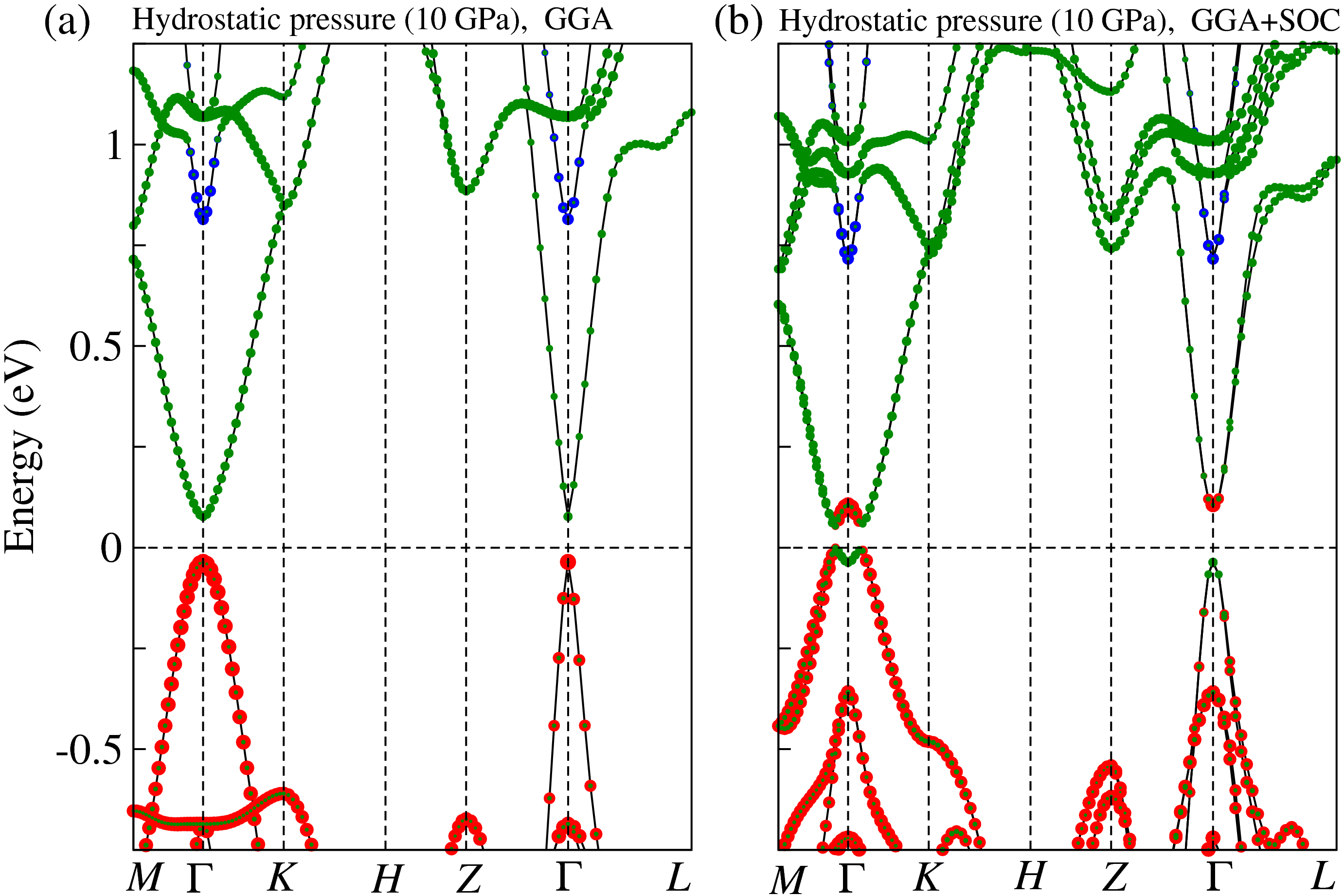}
\end{center}
\caption{(color online)
Bulk band structures for NaBaBi (a) without and (b) with
SOC under a hydrostatic pressure of 10 GPa.
Sizes of blue, red, and green dots represent the weight contributions of
Na $s$, Bi $p$, and Ba $d$ orbitals, respectively. High-symmetry momenta
are implied in Fig. 1(b), and the Fermi energy has been tuned to
zero. Original $s$-$p$ inversion changes to $p$-$d$ inversion.
} \label{press_band}
\end{figure}

As an example, we perform a detailed analysis of the electronic
structures at a hydrostatic pressure of 10 GPa. 
The nontrivial-II phase at other pressures (hydrostatic and uniaxial) 
exhibit the same band order and just different band gaps.
The DFT band structures without and with SOC are
compared in Figs. 5(a) and (b). 
Before SOC is included, the Na $s$ state is pushed up as the second-lowest conduction band compared to the case without pressure, and Ba $d_z^2$ becomes the lowest conduction band.
Then SOC induces band inversion between the Ba $d_z^2$ and Bi $p_z$ states. 
The evolution of the NaBaBi band structures is further illustrated with respect to SOC and pressure. 
To demonstrate the topological feature, we also calculate the surface state on the (001) surface 
using a slab model. The MLWFs here are derived from the Na $s$, Bi $p$, and Ba $d$ atomic-like orbitals.
The corresponding surface band structure is shown in Fig. 6(a).
We can clearly see the Dirac-cone-type TSSs inside the bulk band gap.
Unlike the previous TI phase with $s$-$p$ inversion, however,
this $p$-$d$-inverted TI phase exhibits right-handed spin texture for the upper Dirac cone on the top surface, as shown in Fig. 6(b). Therefore, we obtain different spin textures in a single material, NaBaBi, via band structure engineering.

To understand the origin of the different spin textures in NaBaBi, we also construct the effective Hamiltonian of this system using the invariant theory.\cite{Winkler2003} We focus on the low-energy physics around the $\Gamma$ point, at which the wavevector group is $D_{3h}$. Because the spin is taken into account, we need to consider the double group of $D_{3h}$, which possesses only three 2D irreducible representations (irreps), denoted as $\bar{\Gamma}_{7,8,9}$, which are defined in Appendix A. It turns out that both the $s$ orbital of Na atoms and the $d_{z^2}$ orbital of Ba atoms belong to the $\bar{\Gamma}_7$ irrep, whereas the $p$ orbital of Bi atoms corresponds to the $\bar{\Gamma}_8$ irrep. This suggests that both the $s$-$p$ inversion and $p$-$d$ inversion can be described by the same effective Hamiltonian. Under the basis $\vert \bar{\Gamma}_7,1/2 \rangle,\vert \bar{\Gamma}_7,-1/2 \rangle,\vert \bar{\Gamma}_8,1/2 \rangle,\vert \bar{\Gamma}_8,-1/2 \rangle$, where the $\bar\Gamma_7$ band is for the $s$ or $d$ orbital and the $\bar\Gamma_8$ band is for the $p$ orbital, the effective Hamiltonian can be written as $H_{\alpha} = \epsilon(\bold{k})I_{4 \times 4} + M(\bold{k}) \Gamma_{5} + P_{\alpha} k_z \Gamma_{45} + Q_{\alpha}(k_x\Gamma_{25} - k_y\Gamma_{15})$, where $\epsilon_\alpha(\bold{k}) = \epsilon_{\alpha,0} + \epsilon_{\alpha,1} k^2_{||} + \epsilon_{\alpha,2} k^2_z$, and $M_\alpha(\bold{k}) = M_{\alpha,0} + M_{\alpha,1} k^2_{||} + M_{\alpha,2}k^2_z$. Here $\alpha=s$ or $d$ when the $\bar{\Gamma}_7$ states are given by the $s$ or $d$ orbitals, respectively. $\Gamma_i$ and $\Gamma_{ij}$ are $\Gamma$ matrices, and $\epsilon$, $M$, $P$, and $Q$ are material-dependent parameters. The details of the construction of the model are given in Appendix A.

For comparison with the $ab~initio$ calculations, we choose an open boundary condition along the $z$ direction and calculate the surface states and the corresponding spin texture on a finite slab using the above effective Hamiltonian. We find that the spin texture is right-handed for $\frac{P_\alpha}{Q_\alpha}<0$ and left-handed for $\frac{P_\alpha}{Q_\alpha}>0$, which is shown in detail in the appendices. A comparison of this with the $ab~initio$ calculations suggests $\frac{P_s}{Q_s}>0$ for $s$-$p$ inversion and $\frac{P_d}{Q_d}<0$ for $p$-$d$ inversion. The parameters $P_\alpha$ and $Q_\alpha$ are related to microscopic wave functions at the $\Gamma$ point. $P_\alpha=-a\frac{\hbar}{m}\langle \alpha|\partial_z|p_z\rangle$, and $Q_\alpha=-b\frac{\hbar}{m}\langle \alpha|\partial_x|p_x\rangle$, where $a$ and $b$ are coefficients for the coupling between $|p_z,\sigma\rangle$ and $|p_{\pm},\sigma\rangle$ of the $p$ orbitals due to SOC, as shown in Appendix B. The different crystal environments for the $s$ orbitals in Na atoms and the $d$ orbitals in Ba atoms should be responsible for the opposite signs of $\frac{P_s}{Q_s}$ and $\frac{P_d}{Q_d}$. In the appendices, we also discuss the relationship between the spin texture and mirror Chern number for the $s$-$p$ and $p$-$d$ inversions. This relationship is similar to that for the opposite spin textures in HgTe and HgS.\cite{HgS}

\begin{figure}[htbp]
\begin{center}
\includegraphics[width=0.45\textwidth]{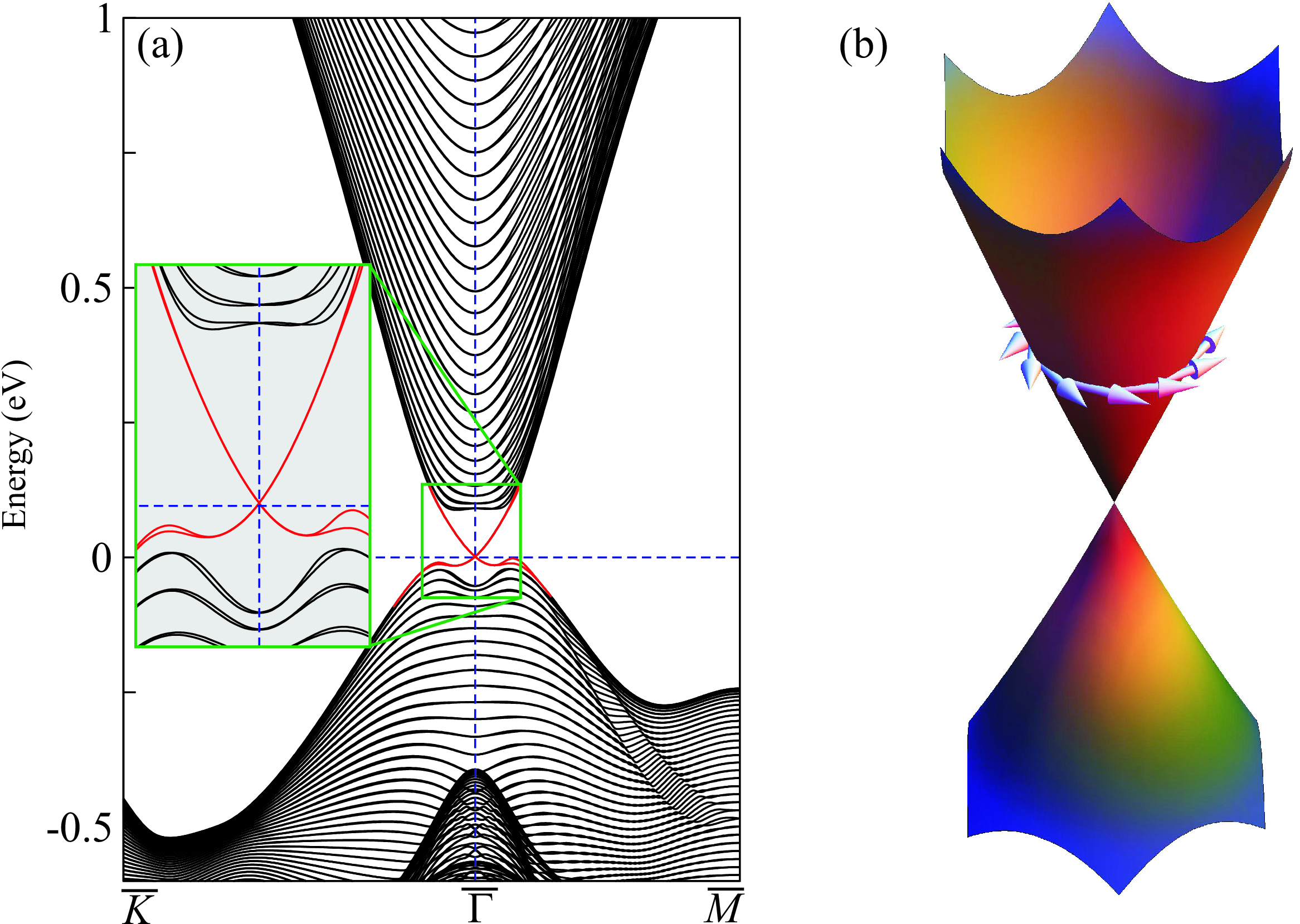}
\end{center}
\caption{(color online)
(a) TB surface band structure on (001) surface 
with $p$-$d$ band inversion under hydrostatic pressure
of 10 GPa. Inset shows local bands around Fermi energy at $\Gamma$
point. (b) Corresponding surface
spin helical Dirac cone. Original left-handed spin texture
in $p$-$s$ inversion phase changes to right-handed.
} \label{press_surface}
\end{figure}

\section{summary}

In conclusion, inspired by the 3D topological Dirac semimetal Na$_3$Bi,
we theoretically propose another
topologically nontrivial state in NaBaBi. Two distinct
TI states with different surface spin textures are
achieved via band engineering. The native
TI phase, with $s$-$p$ band inversion, exhibits left-handed
spin texture in the Dirac-type surface states. Under external pressure, 
the other nontrivial state, which has $p$-$d$ inversion
 and exhibits right-handed helical spin texture, can be induced.
 This is further explained by the effective Hamiltonians constructed, in which 
 $s$-$p$ and $p$-$d$ inverted TI phases exhibit opposite 
 mirror Chern numbers and thus opposite spin helicity.
Because the $d$ orbitals participate
in the band inversion, it is possible to bring the
 correlation effect into the TI phase in this compound.

\begin{acknowledgments}
Thanks to S. Yunoki and the RIKEN
Integrated Cluster of Clusters (RICC) for
computational resources. B.Y. and C.F. acknowledge funding support by the ARCHES award from the Federal German Ministry for Education and Research (BMBF) and the ERC Advanced Grant (291472), respectively.
\end{acknowledgments}

\appendix
 
\section{Construction of effective Hamiltonian by symmetry}

In this section, we will show how to construct the effective Hamiltonian on the basis of the crystal symmetry. The point group at the $\Gamma$ point in momentum space for NaBaBi is $D_{3h}$. The character table for the $D_{3h}$ double group is shown in Table \ref{chracter:d3h}. To distinguish the $\Gamma$ matrices that we will use later, we use $\bar{\Gamma}$ to denote different irreducible representations (irreps). Only double irreps can be used to describe electron systems with spin. Thus, we consider only the $\bar{\Gamma}_{7,8,9}$ irreps. The basis functions for the $\bar{\Gamma}_{7,8,9}$ irreps are denoted as $\vert \bar{\Gamma}_7\rangle = \vert  j=1/2,m_j = \pm 1/2\rangle$, $\vert \bar{\Gamma}_8\rangle = z\vert  j=1/2,m_j = \pm 1/2\rangle$, and $\vert \bar{\Gamma}_9\rangle = \vert j=3/2,m_j = \pm 3/2\rangle$, respectively.

\begin{table}[htb]
  \begin{minipage}[t]{1\linewidth}
      \caption{Character table for the double group of $D_{3h}$. }
\label{chracter:d3h}
\hspace{0cm}
\begin{tabular}{cccccccccc}
    \hline\hline
    $D_{3h}$ \hfill & E \hfill & $\bar{E}$ \hfill & $\sigma_h\&\bar{\sigma}_h$ & $2C_3$ \hfill & $2\bar{C}_3$ \hfill &  $2S_3$ \hfill & $2\bar{S}_3$ \hfill & $3C'_2\&3\bar{C}'_2$ \hfill & $3\sigma_v\&3\bar{\sigma}_v$\\
    \hline
    $\bar{\Gamma}_1$&1&1&1&1&1&1&1&1&1 \\
    $\bar{\Gamma}_2$&1&1&1&1&1&1&1&-1&-1 \\
    $\bar{\Gamma}_3$&1&1&-1&1&1&-1&-1&1&-1\\
    $\bar{\Gamma}_4$&1&1&-1&1&1&-1&-1&-1&1\\
    $\bar{\Gamma}_5$&2&2&-2&-1&-1&1&1&0&0\\
    $\bar{\Gamma}_6$&2&2&2&-1&-1&-1&-1&0&0\\
    $\bar{\Gamma}_7$&2&-2&0&1&-1&$\sqrt{3}$&$-\sqrt{3}$&0&0\\
    $\bar{\Gamma}_8$&2&-2&0&1&-1&$-\sqrt{3}$&$\sqrt{3}$&0&0\\
    $\bar{\Gamma}_9$&2&-2&0&-2&2&0&0&0&0  \\
    \hline
  \end{tabular}%
  \end{minipage}
\end{table}

The orbitals of NaBaBi near the Fermi energy are  $s$ orbital of Na atoms, $p$ orbital of Bi atoms, and $d$ orbital of Ba atoms. From a first-principle calculation, we find that the main component of the $p$ ($d$) orbital of Bi (Ba) atoms is $p_z$ ($d_{z^2}$) orbital. Next, we will determine the irreps to which the $s$, $p_z$, and $d_{z^2}$ orbitals belong.

To distinguish irreps $\bar{\Gamma}_{7,8,9}$, we need to consider only the characters under $C_3$ and $S_3$ operations. $C_3 \vert  s(d_{z^2},p_z), \uparrow \rangle =e^{-i\pi/3} \vert  s(d_{z^2},p_z), \uparrow \rangle$. $C_3 \vert  s(d_{z^2},p_z), \downarrow \rangle =e^{i\pi/3} \vert  s(d_{z^2},p_z), \downarrow \rangle$. All the characters under $C_3$ rotation for the $s$, $p_z$, and $d_{z^2}$ orbitals are $\chi = e^{-i\pi/3} + e^{i\pi/3} = 1$.

For the $S_3$ operation, $S_3 \vert  s(d_{z^2}), \uparrow \rangle = i e^{-i\pi/3} \vert  s(d_{z^2}), \uparrow \rangle$, and $S_3 \vert  s(d_{z^2}), \downarrow \rangle = -i e^{i\pi/3} \vert  s(d_{z^2}), \downarrow \rangle$. The character is $\chi =ie^{-i\pi/3}-ie^{i\pi/3} = \sqrt{3}$. Thus, both the $s$ and $d_{z^2}$ orbitals belong to irrep $\bar{\Gamma}_7$. For the $p_z$ orbital, however, $S_3 \vert p_z,\uparrow \rangle = -i e^{-i\pi/3} \vert p_z,\uparrow \rangle$, and $S_3 \vert p_z,\downarrow \rangle = i e^{i\pi/3} \vert p_z,\downarrow \rangle$. The character is $\chi =-ie^{-i\pi/3}+ie^{i\pi/3} = -\sqrt{3}$. Thus, the $p_z$ orbital belongs to irrep $\bar{\Gamma}_8$.

As shown in Table \ref{tab:d3h}, the $s$ and $d_{z^2}$ orbitals belong to irrep $\bar{\Gamma}_7$, whereas the $p_z$ orbital belongs to irrep $\bar{\Gamma}_8$.

\begin{table}[htb]
  \centering
  \begin{minipage}[t]{0.8\linewidth}
      \caption{ Character table for $s$, $p_z$, and $d_{z^2}$ orbitals with spin.  }
\label{tab:d3h}
\hspace{0cm}
\begin{tabular}{ccccc}
    \hline\hline
    Orbitals with spin \hfill & $2C_3$ \hfill &$2S_3$ & Irreps\\
    \hline
    $s\uparrow$, $s\downarrow$ & 1 & $\sqrt{3}$ & $\bar{\Gamma}_7$\\
    $p_z\uparrow$, $p_z\downarrow$ & 1 & $-\sqrt{3}$ & $\bar{\Gamma}_8$\\
    $d_{z^2}\uparrow$, $d_{z^2}\downarrow$ & 1 & $\sqrt{3}$ & $\bar{\Gamma}_7$\\
    \hline
  \end{tabular}%
  \end{minipage}
\end{table}

After identifying the irreps for each orbital near the Fermi energy, we use the representation theory to construct the effective model for NaBaBi near the $\Gamma$ point in momentum space. Basically, we follow the procedure in Ref. \onlinecite{liu2010}.

The effective Hamiltonian we will construct has the following basis: $\xi = (\vert  \bar{\Gamma}_7,1/2 \rangle,\vert  \bar{\Gamma}_7,-1/2 \rangle,\vert  \bar{\Gamma}_8,1/2 \rangle,\vert  \bar{\Gamma}_8,-1/2 \rangle)^T$. The transformation matrices of the symmetry operations in $D_{3h}$ group under this basis are
\begin{itemize}
\item Time reversal operation: $T = diag[-i\sigma_2K,-i\sigma_2K] = -\tau_3 \otimes i\sigma_2K$;
\item Mirror operation along the $z$ direction: $\mathcal{M}_z = diag[i\sigma_3,-i\sigma_3] = \tau_3 \otimes i\sigma_3$;
\item $C_3$ operation: $C_3 = diag[e^{-i\frac{\pi}{3}},e^{i\frac{\pi}{3}},e^{-i\frac{\pi}{3}},e^{i\frac{\pi}{3}}]$;
\item $C_{2,x}$ operation along the $x$ direction: $C_{2,x} = diag[-i\sigma_1,i\sigma_1] = -\tau_3 \otimes i\sigma_1$;
\item Mirror operation along the $y$ direction: $\mathcal{M}_y = diag[i\sigma_2,i\sigma_2] = \tau_0 \otimes i\sigma_2$,
\end{itemize}
where $\sigma_i$ and $\tau_i$ are the Pauli matrices in spin and band space, respectively. Here we select a particular gauge such that $T = i\sigma_2K$ in spin-up and spin-down space, where $K$ represents complex conjugation. We set $\vert  \bar{\Gamma}_7,\pm 1/2 \rangle = i \vert  s(d_{z^2}) ,\uparrow(\downarrow)\rangle$ to obtain $T = -\tau_3 \otimes i\sigma_2K$. By using this convention, we make all the parameters real in the Hamiltonian that we will derive.

We start with 5 Dirac $\Gamma$ matrices on the basis mentioned above, which are defined as
\begin{eqnarray}
    &&\Gamma_1=\tau_1 \otimes \sigma_1,\qquad
    \Gamma_2=\tau_1 \otimes \sigma_2,\qquad
    \Gamma_3=\tau_1 \otimes \sigma_3,\nonumber\\
    &&\Gamma_4=\tau_2 \otimes 1,\qquad
    \Gamma_5=\tau_3 \otimes 1.
    \label{GamMat_1}
\end{eqnarray}
They satisfy Clifford algebra: $\{\Gamma_a,\Gamma_b\}=2\delta_{ab}$. The other 10 $\Gamma$ matrices are constructed by $\Gamma_{ab}=[\Gamma_a,\Gamma_b]/2i$ and are expressed explicitly as
\begin{eqnarray}
    &&\Gamma_{ij}=[\tau_1 \otimes\sigma_i,\tau_1 \otimes \sigma_j]/2i
    =1 \otimes \varepsilon_{ijk}\sigma_k,\\
    &&\Gamma_{i4}=[\tau_1\otimes\sigma_i,\tau_2 \otimes 1]/2i
    =\tau_3\otimes\sigma_i,\\
    &&\Gamma_{i5}=[\tau_1\otimes\sigma_i,\tau_3\otimes 1]/2i=-\tau_2\otimes\sigma_i,\\
    &&\Gamma_{45}=[\tau_2\otimes 1,\tau_3 \otimes 1]/2i=\tau_1\otimes 1,
    \label{GamMat_2}
\end{eqnarray}
where $i,j=1,2,3$.

Under the time-reversal transformation matrix, the $\Gamma$ matrices satisfy
\begin{eqnarray}
    &&T\Gamma_iT^{-1} = \Gamma_i,\qquad i=1,2,3,4,5;\\
    &&T\Gamma_{ij}T^{-1}= - \Gamma_{ij}, \qquad i,j = 1,2,3, \qquad  i\neq j;\\
    &&T\Gamma_{i4}T^{-1}= - \Gamma_{i4}, \qquad i = 1,2,3;\\
    &&T\Gamma_{i5}T^{-1}= - \Gamma_{i5}, \qquad i = 1,2,3;\\
    &&T\Gamma_{45}T^{-1}= - \Gamma_{45}.
\end{eqnarray}

Using the $\mathcal{M}_z$, $C_{2,x}$, and $\mathcal{M}_y$ transformation matrices, we have
\begin{widetext}
\begin{eqnarray}
\nonumber \mathcal{M}_z:     &&\mathcal{M}_z\Gamma_{1,2,5,12,34,15,25}\mathcal{M}_z^{-1} = \Gamma_{1,2,5,12,34,15,25},\\
                    &&\mathcal{M}_z\Gamma_{3,4,23,31,14,24,35,45}\mathcal{M}_z^{-1} = -\Gamma_{3,4,23,31,14,24,35,45}, \\
\nonumber C_{2,x}:    &&C_{2,x}\Gamma_{2,3,5,23,14,25,35}C_{2,x}^{-1} = \Gamma_{2,3,5,23,14,25,35},\\
                            &&C_{2,x}\Gamma_{1,4,31,12,24,34,15,45}C_{2,x}^{-1} = -\Gamma_{1,4,31,12,24,34,15,45},\\
 \nonumber \mathcal{M}_y:     &&\mathcal{M}_y\Gamma_{2,4,5,31,24,25,45}\mathcal{M}_y^{-1} = \Gamma_{2,4,5,31,24,25,45},\\
                    &&\mathcal{M}_y\Gamma_{1,3,12,23,14,34,15,35}\mathcal{M}_y^{-1} = - \Gamma_{1,3,12,23,14,34,15,35}.
\end{eqnarray}
\end{widetext}

Under $C_3$ rotation symmetry, the $\Gamma$ matrices satisfy $\Gamma'_{\theta} = e^{i\Sigma/2\theta} \Gamma e^{-i\Sigma/2\theta}$, with $\Sigma = 1 \otimes \sigma_z$. Furthermore, $\frac{d\Gamma'_{\theta}}{d\theta}  = \frac{i}{2}[\Sigma,\Gamma'_{\theta}]$. Thus, the commutation relations between $\Sigma$ and $\Gamma$ are important here, and are listed as follows:
\begin{eqnarray}
    &&[\Sigma,\Gamma_1]=2i\Gamma_2,\qquad[\Sigma,\Gamma_2]=-2i\Gamma_1,\\
    &&[\Sigma,\Gamma_3]=[\Sigma,\Gamma_4]=[\Sigma,\Gamma_5]=0,\\
    &&[\Sigma,\Gamma_{12}]=0,\qquad [\Sigma,\Gamma_{34}]=0,\\
    &&[\Sigma,\Gamma_{31}]=-2i\Gamma_{23},\qquad
    [\Sigma,\Gamma_{23}]=2i\Gamma_{31},\\
    &&[\Sigma,\Gamma_{14}]=2i\Gamma_{24},\qquad
        [\Sigma,\Gamma_{24}]=-2i\Gamma_{14},\\
    &&[\Sigma,\Gamma_{15}]=2i\Gamma_{25},\qquad
    [\Sigma,\Gamma_{25}]=-2i\Gamma_{15},\\
    &&[\Sigma,\Gamma_{35}]=0,\qquad [\Sigma,\Gamma_{45}]=0.
    \label{eq:commutation}
\end{eqnarray}
Using these commutation relations, we can easily solve for $\Gamma'_{\theta}$, which are expressed as
\begin{eqnarray}
    &&\Gamma'_1(\theta)=\Gamma_1\cos\theta+\Gamma_2\sin\theta,\nonumber\\
    &&\Gamma'_2(\theta)=\Gamma_1\sin\theta-\Gamma_2\cos\theta,\\
    &&\Gamma'_{23}(\theta)=\Gamma_{23}\cos\theta+\Gamma_{31}\sin\theta,\nonumber\\
    &&\Gamma'_{31}(\theta)=\Gamma_{31}\cos\theta-\Gamma_{23}\sin\theta,\\
    &&\Gamma'_{14}(\theta)=\Gamma_{14}\cos\theta+\Gamma_{24}\sin\theta,\nonumber\\
    &&\Gamma'_{24}(\theta)=\Gamma_{14}\sin\theta-\Gamma_{24}\cos\theta,\\
    &&\Gamma'_{15}(\theta)=\Gamma_{15}\cos\theta+\Gamma_{25}\sin\theta,\nonumber\\
    &&\Gamma'_{25}(\theta)=\Gamma_{15}\sin\theta-\Gamma_{25}\cos\theta,\\
    &&\Gamma'_3(\theta)=\Gamma_3,\qquad\Gamma'_4(\theta)=\Gamma_4, \nonumber\\
    &&\Gamma'_5(\theta)=\Gamma_5,\qquad\Gamma'_{34}=\Gamma_{34},
    \qquad\Gamma'_{12}=\Gamma_{12},\nonumber\\
    &&\Gamma'_{35}=\Gamma_{35},\qquad \Gamma'_{45}=\Gamma_{45}.
    \label{eq:expGamma1}
\end{eqnarray}
$\Gamma$ matrix pairs $[\Gamma_1,\Gamma_2]$,  $[\Gamma_{23},\Gamma_{31}]$, $[\Gamma_{14},\Gamma_{24}]$, and $[\Gamma_{15},\Gamma_{25}]$ transform as the $x$,$y$ components of a vector under $C_3$ rotation. The other seven $\Gamma$ matrices transform as scalars or pseudo-scalars under $C_3$ rotation.

Because $\bar{\Gamma}_7 \otimes \bar{\Gamma}_8 = \bar{\Gamma}_3 \oplus \bar{\Gamma}_4\oplus \bar{\Gamma}_5${}, here we consider only irreps $\bar{\Gamma}_3$, $\bar{\Gamma}_4$, and $\bar{\Gamma}_5$ for the effective model. The character table for these three irreps is given in Table \ref{tab:gamma}.

\begin{table}[htb]
  \centering
  \begin{minipage}[t]{0.8\linewidth}
      \caption{ Character table of $\Gamma$ matrices and polynomials of
      the momentum ${\bold{k}}$.  }
\label{tab:gamma}
\hspace{0cm}
\begin{tabular}{ccccc}
    \hline\hline
    Representation\hfill & Basis functions\hfill &T\\
    \hline
    $\bar{\Gamma}_3$& $\Gamma_3$            & +  \\
                    & $\Gamma_{35}$         & -  \\
    $\bar{\Gamma}_4$  & $k_z$               & -  \\
                & $\Gamma_4$                & +  \\
                & $\Gamma_{45}$             & -  \\
    $\bar{\Gamma}_5$  & $(k_x,k_y)$         & -  \\
                & $(k^2_x-k^2_y,k_xk_y)$    & +  \\
                & $(\Gamma_1,\Gamma_2)$     & +  \\
                & $(\Gamma_{15},\Gamma_{25})$ & -  \\
    \hline
  \end{tabular}%
  \end{minipage}
\end{table}

Under $C_3$ rotation, we have two rotationally invariant combinations of $\Gamma$ matrices and $\bold{k}$ that preserve time-reversal symmetry. They are $k'_x \Gamma'_{25} - k'_y\Gamma'_{15} = k_x \Gamma_{25} - k_y\Gamma_{15}$ and  $k'_x \Gamma'_{15} + k'_y\Gamma'_{25} = k_x \Gamma_{15} + k_y\Gamma_{25}$. However, under the $\mathcal{M}_y$ operation, $\mathcal{M}_y \Gamma'_{15} \mathcal{M}^{-1}_y= -\Gamma'_{15}$, and $\mathcal{M}_y \Gamma'_{25} \mathcal{M}^{-1}_y= +\Gamma'_{25}$; $\mathcal{M}_y k_x = k_x$, and $\mathcal{M}_y k_y = -k_y$. The effective Hamiltonian needs to preserve the mirror symmetry $\mathcal{M}_y$ and time-reversal symmetry simultaneously. Therefore, the qualified term reads $k_x \Gamma_{25} - k_y\Gamma_{15}$. Similarly, we obtain another term, $2k_xk_y\Gamma_1 + (k^2_x-k^2_y)\Gamma_2$. Thus, the effective Hamiltonian on the new basis, $\xi = (\vert  \bar{\Gamma}_7,1/2 \rangle,\vert  \bar{\Gamma}_7,-1/2 \rangle,\vert  \bar{\Gamma}_8,1/2 \rangle,\vert  \bar{\Gamma}_8,-1/2 \rangle)^T$, up to the second order of $\bold{k}$, is expressed as $H = \epsilon(\bold{k})I_{4 \times 4} + M(\bold{k}) \Gamma_{5} + Pk_z \Gamma_{45} + Q(k_x\Gamma_{25} - k_y\Gamma_{15})+ Q'(2k_xk_y\Gamma_1 + (k^2_x-k^2_y)\Gamma_2)$. Explicitly, the effective Hamiltonian can be written as
\begin{widetext}
\begin{eqnarray}
 &&H_{eff}= \left(
	\begin{array}{cccccc}
          M_7(\bold{k})&0&Pk_z&Qk_-+Q'(2k_xk_y - i(k^2_x-k^2_y))\\
          &M_7(\bold{k})&-Qk_+ +Q'(2k_xk_y + i(k^2_x-k^2_y))&Pk_z\\
            &  h.c.&   M_8(\bold{k})&0\\
            &   &      &M_8(\bold{k})
    \end{array},
\label{eq:eff_Ham}
	\right)
\end{eqnarray}
\end{widetext}
where
\begin{eqnarray}
\nonumber && \epsilon(\bold{k}) = \epsilon_0 + \epsilon_1 k^2_{\vert \vert } + \epsilon_2 k^2_z, \\
\nonumber && M(\bold{k}) = M_0 + M_1 k^2_{\vert \vert } + M_2k^2_z,\\
\nonumber && M_7(\bold{k}) = \epsilon(\bold{k}) + M(\bold{k}),\\
&& M_8(\bold{k}) = \epsilon(\bold{k}) - M(\bold{k}),
\end{eqnarray}
and $\epsilon_i$, $M_i$, $P$, $P'$, and $Q$ are real. Because $\epsilon(\bold{k})$, $M(\bold{k})$, $I_{4\times 4}$, and $\Gamma_5$ belong to the $\bar{\Gamma}_1$ irrep and they are all even under time-reversal symmetry, combinations of these terms, $\epsilon(\bold{k})I_{4 \times 4}$ and $M(\bold{k}) \Gamma_{5}$, are invariant under all the operations of the $D_{3h}$ point group.  Thus, we also need to include them in the effective Hamiltonian. This Hamiltonian is used to explain the spin textures for both the $s$-$p$ and $d$-$p$ inversions in NaBaBi.

\section{Spin helicity and mirror Chern number}

\begin{figure}[tb]
	\includegraphics[width = 0.9\columnwidth,angle=0]{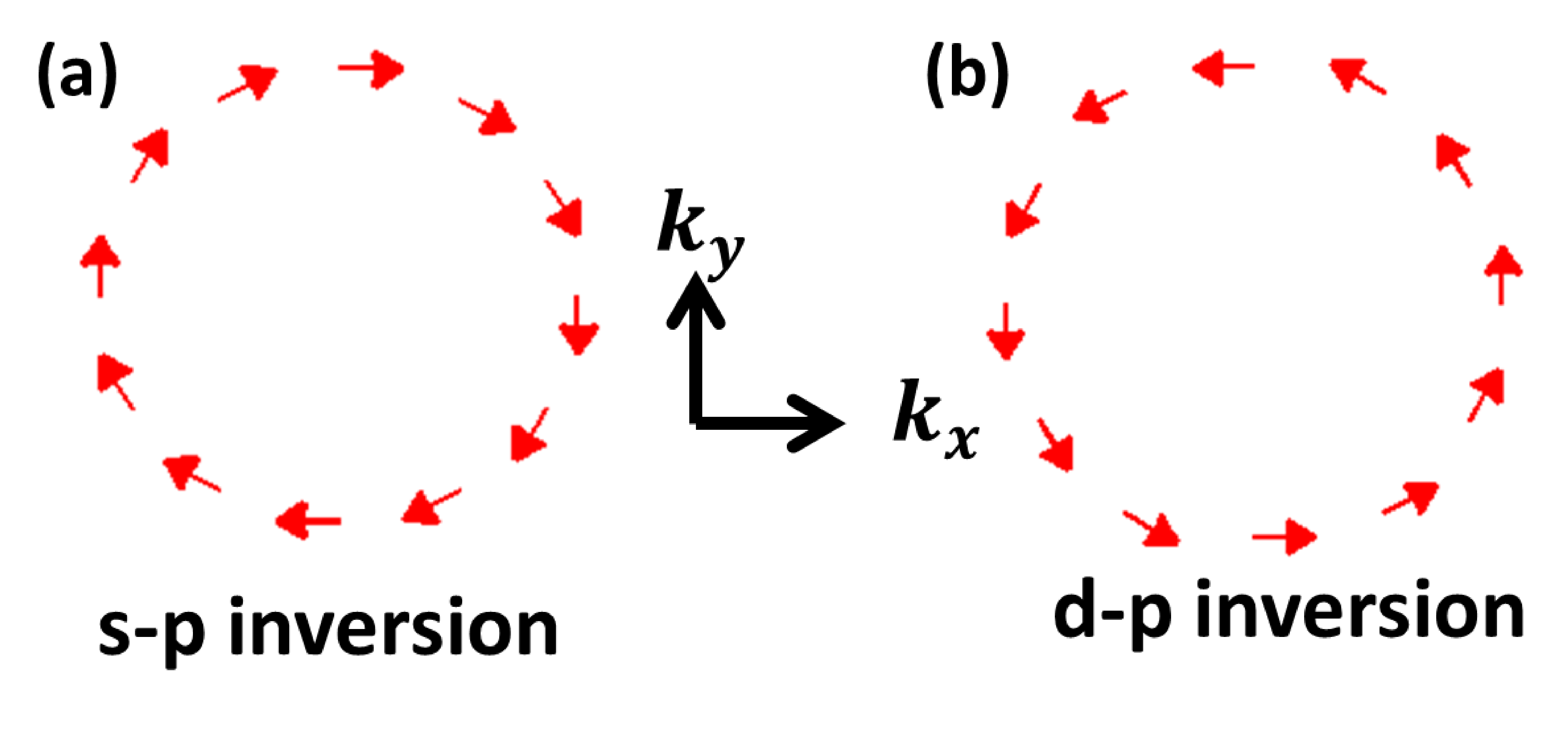}
  \caption{
   (Color online). Spin textures on the Fermi arc with energy $E_F = 0.1$ eV (compared with the Dirac point energy) on the upper Dirac cone of the top surface of NaBaBi. Panels (a) and (b) correspond to $s$-$p$ and $d$-$p$ inversion in NaBaBi, which have left-handed and right-handed spin textures, respectively, on the upper Dirac cone of the top surface of NaBaBi.
  }
    \label{fig1}
\end{figure}

We consider a slab described by the effective Hamiltonian in Eq. (\ref{eq:eff_Ham}) with the open boundary along the $k_z$ direction. The spin texture is calculated numerically with appropriate parameters for qualitative agreement with the result obtained by the first-principle methods. The parameters are listed in Table \ref{tab1}. In Fig. \ref{fig1}, we show the Fermi arc and spin texture on the upper Dirac cone of the top surface of NaBaBi on the $k_xk_y$ plane. The Fermi arc is a circle, and the directions of the spins are tangential to the circle. The spin texture is right-handed (left-handed) for NaBaBi with $d$-$p$ inversion ($s$-$p$ inversion), which is consistent with the results of the first-principle calculations. The key point for the opposite spin textures is found to be the sign of $P/Q$. For $P/Q>0$, the spin texture is left-handed, whereas for $P/Q<0$, it is right-handed.

\begin{widetext}
\begin{center}
\begin{table}[htb]
  \centering
  \begin{minipage}[t]{1\linewidth}
  \caption{ Parameters for NaBaBi. For $s$-$p$ inversion, $\epsilon_1$ = 0.05 eV $\cdot$ \AA,~ $Q = 0.5$ eV $\cdot$ \AA, and~ $Q' = -0.2$ eV $\cdot$ \AA$^2$; for $d$-$p$ inversion, $\epsilon_1 = -0.05$ eV $\cdot$ \AA,~ $Q = -0.5$ eV $\cdot$ \AA, and~ $Q' = 0.2$ eV $\cdot$ \AA$^2$.}
\label{tab1}
\hspace{-1cm}
\begin{tabular}[c]{ccccccccc}\hline\hline
$\epsilon_0$[eV]&$\epsilon_1$[eV $\cdot$ \AA$^2$]&$\epsilon_2$[eV $\cdot$ \AA$^2$]&$M_0$[eV]&$M_1$[eV $\cdot$ \AA$^2$]&$M_2$[eV $\cdot$ \AA$^2$]&$P$[eV $\cdot$ \AA]&$Q$[eV $\cdot$ \AA]&$Q'$[eV $\cdot$ \AA$^2$] \\
0&$\pm$ 0.05&0&-0.5 & 5 &5&1&$\pm$ 0.5 & $\pm$ 0.2\\\hline\hline\hline
\label{para2}
\end{tabular}
  \end{minipage}
\end{table}
\end{center}
\end{widetext}

This can be understand from the mirror symmetry along the $y$ direction. Next, we hope to unveil the physics behind the relationship between $P/Q$ and the spin texture. Because the system preserves mirror symmetry with respect to the $k_y = 0$ plane, the Hamiltonian commutes with the mirror operator $\mathcal{M}_y$: $\mathcal{M}_y H(k_x,0,k_z)\mathcal{M}^{-1}_y = H(k_x,0,k_z)$. Therefore, the Hamiltonian and the mirror operator can be diagonalized simultaneously. The mirror operator along the $y$ direction is given by $\mathcal{M}_y = \tau_0 \otimes i\sigma_2$. In the $+i$ mirror parity subspace, the common eigen-wavefunctions are $\Psi(+i)_1 = \frac{1}{\sqrt{2}}(1,i,0,0)^T$ and $\Psi(+i)_2 = \frac{1}{\sqrt{2}}(0,0,1,i)^T$. On these two basis functions, the projected Hamiltonian reads
\begin{widetext}
\begin{eqnarray}
&&H_{+i}=\left(
	\begin{array}{cc}
          M_7&P(k_z+i\frac{Q}{P}k_x)+Q'k^2_x\\
          P(k_z-i\frac{Q}{P}k_x)+Q'k^2_x&M_8
    \end{array}
\label{eq:mirror_i1}
	\right)
\end{eqnarray},
\end{widetext}
where $M_{7}(\bold{k}) = \epsilon(\bold{k}) + M(\bold{k})$, and $M_{8}(\bold{k}) = \epsilon(\bold{k}) - M(\bold{k})$. $Q'$ is much smaller than $P$ and $Q$ and is neglected in the analysis below.

The Hamiltonian in the $+i$ mirror parity subspace has a nonzero Chern number, which can be calculated from $n = \int_{BZ} \frac{d^2k}{4\pi} \hat{\mathbf{d}} \cdot (\frac{\partial \hat{\mathbf{d}}}{\partial k_z} \times \frac{\partial \hat{\mathbf{d}}}{\partial k_x})$, where $\hat{\mathbf{d}} = \frac{1}{\vert \mathbf{d}\vert }(d_x, d_y, d_z)$ when we write $H_{+i}$ as $H_{+i} = \epsilon(\bold{k})+\bold{d}\cdot \bold{\sigma}$.\cite{qi2006} Because the Chern number $n_{-i}$ for the effective Hamiltonian in the $-i$ mirror parity subspace always takes the opposite value because of time-reversal symmetry, $n_{-i} = -n_{+i}$, the total Chern number is zero, and only the mirror Chern number, defined as $n_M \equiv \frac{n_{+i}-n_{-i}}{2} = n_{+i}$\cite{hsieh2012}, can be nonzero. For $P/Q>0$, $n_M = -1$, whereas for $P/Q<0$, $n_M = 1$. The mirror Chern number is directly related to the spin texture of the surface states according to the bulk--edge correspondence. Now let us consider that the open boundary along the $z$ direction and the mirror symmetry along the $y$ direction still exist for $k_y = 0$.  For the $+i$ mirror parity subspace and $n_{+i} = 1$, there is a chiral edge mode with positive velocity along the $k_x$ direction. Because the mirror parity $+i$ subspace corresponds to spin up along the $y$ direction, this chiral mode should have spin up along the $y$ direction. The spin-down state along the $y$ direction is in the $-i$ mirror parity subspace and thus possesses negative velocity.\cite{HgS} Therefore, the entire system has a right-handed spin texture for $P/Q<0$. When $P/Q>0$, the mirror Chern number is $-1$, and the spin texture is left-handed.

Next, we will explain what determines the opposite signs of $P/Q$ for $s$-$p$ and $d$-$p$ inversion. Without losing generality, we neglect the off-diagonal quadratic term for simplicity. We can write the basis functions for irreps $\bar{\Gamma}_7$ and $\bar{\Gamma}_8$ in terms of the $s$, $p_z$, and $d_{z^2}$ orbitals. In the $+i$ mirror parity subspace, there are two ways to express the basis functions for irrep $\bar{\Gamma}_7$, $\frac{1}{\sqrt{2}}(\vert \bar{\Gamma}_7,1/2\rangle_z+i \vert \bar{\Gamma}_7,-1/2\rangle_z)$, which are $\Psi(s)_{+i} = i\frac{1}{\sqrt{2}}(\vert  s,\uparrow_z\rangle+i\vert  s,\downarrow_z\rangle)$ and $\Psi(d_{z^2})_{+i} = i \frac{1}{\sqrt{2}} (\vert  d_{z^2},\uparrow_z\rangle + i\vert  d_{z^2},\downarrow_z\rangle)$, where the spins are along the $z$ direction. The basis function for irrep $\bar{\Gamma}_8$ in the mirror parity $\lambda = +i$ subspace is $\Psi(p_z)_{+i} = \frac{1}{\sqrt{2}}(\vert \bar{\Gamma}_8,1/2\rangle_z+i \vert \bar{\Gamma}_8,1/2\rangle_z)$.

Because of SOC, $\vert  p_z, \uparrow_z\rangle$ can be mixed with the $\vert  (p_x+ip_y), \downarrow_z\rangle$ state. Similarly, the $\vert  p_z,\downarrow_z\rangle$ state can be mixed with the $\vert  (p_x-ip_y), \uparrow_z\rangle$ state. The basis functions for $\bar{\Gamma}_8$ irrep, written in terms of the $p$ orbitals, read
\begin{eqnarray}
\nonumber && \vert \bar{\Gamma}_8,1/2\rangle_z = a \vert  p_z,\uparrow_z\rangle -b\vert  (p_x+ip_y), \downarrow_z\rangle, \\
&&\vert \bar{\Gamma}_8,-1/2\rangle_z = a \vert  p_z,\downarrow_z\rangle +b\vert  (p_x-ip_y),\uparrow_z\rangle,
\end{eqnarray}
where $a^2 + 2b^2 = 1$. Because the main component of the $p$ orbitals is the $p_z$ orbital, we regard the parameter $b$ as a perturbation. We have $a \gg b$. Here we neglect SOC between the $eg$ and $t2g$ orbitals because of the large energy splitting between them.

To obtain the Hamiltonian in Eq. (\ref{eq:mirror_i1}), we transform the basis functions above to a set of new basis functions, where the spin quantization direction is along the $y$ direction (the normal direction of the mirror plane). For each basis function in the mirror $\lambda = +i$ subspace, we have
\begin{widetext}
\begin{eqnarray}
\nonumber \Psi(s)_{+i} && = i \vert  s, \uparrow_y\rangle,\\
\nonumber \Psi(d)_{+i} && = i \vert  d_{z^2}, \uparrow_y\rangle, \\
 \Psi(p)_{+i} &&= \frac{1}{\sqrt{2}}(\vert \bar{\Gamma}_8,1/2\rangle_z + i\vert \bar{\Gamma}_8,-1/2\rangle_z) = a\vert  p_z, \uparrow_y\rangle + ib \vert  p_x, \uparrow_y\rangle + b\vert  p_y, \downarrow_y\rangle,
\label{eq:basis}
\end{eqnarray}
\end{widetext}
where we have used $\vert \uparrow_y\rangle = \frac{1}{\sqrt{2}}(\vert \uparrow_z\rangle + i\vert \downarrow_z\rangle)$.

Thus, $P$ and $Q$ in Eq. (\ref{eq:mirror_i1}) can be derived in terms of $a$, $b$, and the matrix elements from $\bold{k}\cdot \bold{p}$ theory. \cite{voon2009} The $\bold{k}\cdot \bold{p}$ Hamiltonian reads $H_{\bold{k}\cdot \bold{p}} = \frac{\hbar}{m}\bold{k} \cdot \bold{p} = \frac{\hbar}{m}\bold{k} \cdot (-i\hbar\bold{\nabla}) = -i\frac{\hbar^2}{m}\bold{k} \cdot \bold{\nabla}$, where we replace $\bold{p}$ with $-i\hbar\bold{\nabla}$ to distinguish it from the notation for the $p$ orbitals. For $s$-$p$ inversion, $P = -i\frac{\hbar^2}{m}\langle \Psi(s)_{+i} \vert  \partial_z\vert\Psi(p)_{+i}\rangle  = -\frac{\hbar^2}{m}a\langle s\vert\partial_z \vert p_z\rangle$, and  $iQ = -i\frac{\hbar^2}{m}\langle\Psi(s)_{+i}\vert \partial_x \vert  \Psi(p)_{+i} \rangle  = -ib\frac{\hbar^2}{m}\langle s\vert  \partial_x \vert p_x\rangle$. Recall that the $s$ orbital is from Na atoms and the $p$ orbitals are from Bi atoms. 
Similarly, for $d$-$p$ inversion of NaBaBi, $P = -i\frac{\hbar^2}{m}\langle\Psi(d_{z^2})_{+i} \vert \partial_z \vert\Psi(p)_{+i}\rangle  = -\frac{\hbar^2}{m}a\langle d_{z^2} \vert\partial_z \vert p_z \rangle$, and $iQ = -i\frac{\hbar^2}{m}\langle\Psi(d_{z^2})_{+i}\vert \partial_x \vert\Psi(p)_{+i}\rangle = -i\frac{\hbar^2}{m}b\langle d_{z^2}\vert\partial_x \vert p_x\rangle$. 

Therefore, we have $P/Q = \frac{a\langle s\vert\partial_z \vert p_z\rangle}{b\langle s\vert  \partial_x \vert p_x\rangle}$ for $s$-$p$ inversion, and  $P/Q = \frac{a\langle d_{z^2} \vert\partial_z \vert p_z \rangle}{b\langle d_{z^2}\vert\partial_x \vert p_x\rangle}$ for $d$-$p$ inversion. This suggests that for $s$-$p$ and $d$-$p$ inversion, the matrix elements $ \frac{\langle s\vert\partial_z \vert p_z\rangle}{\langle s\vert  \partial_x \vert p_x\rangle}$ and $\frac{\langle d_{z^2} \vert\partial_z \vert p_z \rangle}{\langle d_{z^2}\vert\partial_x \vert p_x\rangle}$  have opposite signs for NaBaBi.


\end{document}